\documentclass[prl,twocolumn,reprint,amsmath,amssymb,amsfonts,showpacs]{revtex4}

\usepackage{graphicx}
\usepackage{multirow}

\begin{document}

\title{Quantifying mixed-state quantum entanglement by optimal entanglement witness}
\author{S.-S. B. Lee}
\author{H.-S. Sim}
\affiliation{Department of Physics, Korea Advanced Institute of Science and Technology, Daejeon 305-701, Korea}

\date{\today}

\begin{abstract}
We develop an approach of quantifying entanglement in mixed quantum states by the optimal entanglement witness operator. We identify the convex set of mixed states for which a single witness provides the exact value of an entanglement measure, and show that the convexity and symmetries of entanglement or of a target state considerably fix the form of the optimal witness. This greatly reduces difficulty in computing and experimentally determining entanglement measures. As an example, we show how to experimentally quantify bound entanglement in four-qubit noisy Smolin states and three-qubit Greenberger-Horne-Zeilinger (GHZ) entanglement under white noise.
\end{abstract}
\maketitle

{\it Introduction.--} Entanglement is at the center of foundational issues about quantum strangeness such as nonlocality. It is a key resource of quantum technologies~\cite{Plenio_review}.

A long-standing issue is how to detect and quantify entanglement~\cite{Plenio_review,Guhne}.
Entanglement measures have been proposed, to quantify entanglement, to estimate its role in quantum information tasks~\cite{Plenio_review,Guhne},
and to characterize its various aspects of such as multipartite entanglement classes~\cite{Guhne} and correlations in many-body states and phase transitions~\cite{Amico}. Many useful measures $\mathcal{E}$, such as entanglement of formation, concurrence, 3-tangle,  etc., are defined to be computable for pure states $\psi$, and extended to mixed states $\rho$ via the convex roof construction
\begin{equation}
  \label{convroof}
  \mathcal{E}(\rho) = \inf_{\{p_i , \psi_i \}} \sum_i p_i \; \mathcal{E}(|\psi_i \rangle),
\end{equation}
where the infimum is taken over all possible pure-state decompositions $\rho = \sum_i p_i |\psi_i \rangle \langle \psi_i |$ with $\sum_i p_i = 1$, $p_i \ge 0$, and $\langle \psi_i | \psi_i \rangle = 1$. This construction satisfies the requirement~\cite{Plenio_review} that a convex measure $\mathcal{E}$ does not increase under local operations and classical communications.

It is hard to compute and to experimentally determine $\mathcal{E}$ for mixed states. For two qubits, concurrence is computable~\cite{Wootters} and experimentally determinable~\cite{Schmid,Park}. 
For higher dimensional or multipartite cases, however, the computation of $\mathcal{E}(\rho)$ requires an impractical numerical task of exploring all possible pure-state decompositions of $\rho$, except for rare specific states~\cite{Terhal_PRL,Verstraete_4qubit,Wei04,Lohmayer} with very low rank or high symmetry. This difficulty
obstructs quantitative researches on the features of mixed-state entanglement such as nonlocality, fragility under noise, dynamics, and applicability to quantum technologies.

In this work, we develop a quantification approach, based on a physical observable known as the optimal entanglement witness operator~\cite{Terhal,Brandao} whose expectation value provides the exact value of an entanglement measure for a state; the optimality is differently defined in Ref.~\cite{Lewenstein}. We reveal the basic relation (see Theorems) between the optimal witness $X_\rho$ for $\mathcal{E}(\rho)$ and the optimal pure-state decomposition of $\rho$ in Eq.~\eqref{convroof}. It allows one to identify the convex set of mixed states for which a single witness provides the exact value of a measure, and combining with symmetries of entanglement (e.g., SLOCC invariance) and a target state, it considerably fixes the form of the optimal witness; SLOCC stands for stochastic local operations and classical communications. These findings greatly reduce the difficulty of computing and experimentally determining $\mathcal{E}$, and apply to general convex measures. As an example, we show how to experimentally quantify bound entanglement in 4-qubit noisy Smolin state~\cite{Augusiak,Lavoie} and 3-qubit 
GHZ entanglement~\cite{GHZ} under noise such as full-rank white noise.



%
%

\begin{figure*}[t]
  \includegraphics[width=.95\textwidth]{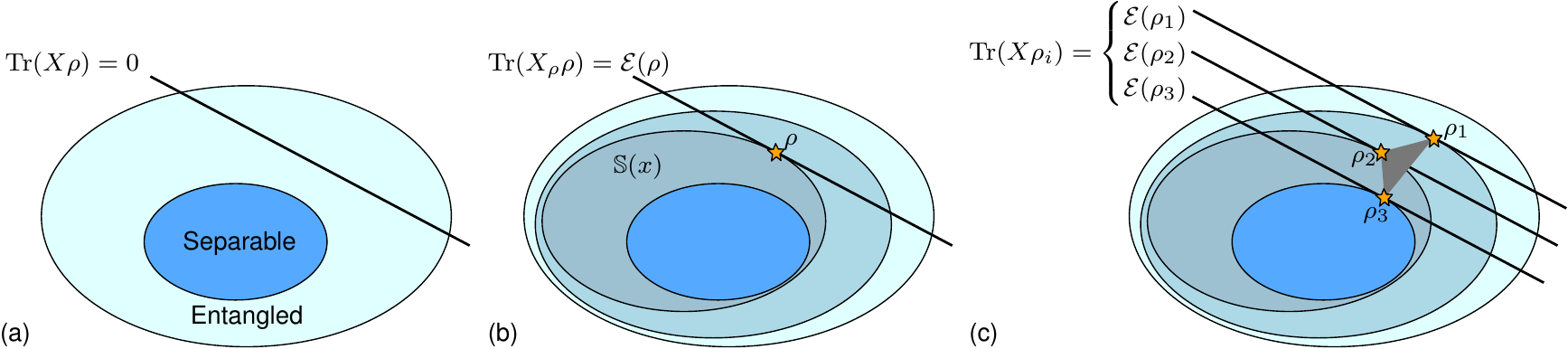}
  \caption{
  (Color Online) Optimal entanglement witness.
  (a) A witness operator $X$ detects entanglement in a quantum state $\rho$, as it divides the set (shade area) of states into two parts with and without separable states by the sign of the expectation value.
  (b) Hierarchy of convex sets $\mathbb{S}(x)$, each consisting of the states $\rho'$ with an entanglement measure $\mathcal{E}(\rho') \le x$; darker for smaller $x$. The witness $X_\rho$ optimal for $\rho$ quantifies entanglement by $\mathcal{E}(\rho) = \textrm{Tr}(X_\rho \rho)$, which corresponds to the support hyperplane (solid line) of $\mathbb{S}(x=\mathcal{E}(\rho))$ at $\rho$.
  (c) When $X$ is optimal for $\{ \rho_i \}$ ($i=1,2,\cdots$), so is it for all states in the convex hull, $\mathrm{conv}(\{ \rho_i \})$, of $\{ \rho_i \}$ (shade triangle).
  For $^\forall \rho \in \mathrm{conv}(\{ \rho_i \})$,
  one computes $\mathcal{E}(\rho)$ by $\textrm{Tr}(X \rho)$,
  without exploring pure-state decompositions of $\rho$.
  }
  \label{fig1}
\end{figure*}

{\it Approach.--} Our approach for entanglement quantification starts with an equivalent expression of $\mathcal{E}(\rho)$~\cite{Brandao,Eisert},
\begin{align}
  \mathcal{E}(\rho) &= \sup_{X \in \mathbb{M}_{\mathcal{H}}} \mathrm{Tr} (X \rho),
  \label{xdef} \\
  \mathbb{M}_{\mathcal{H}} &= \{ X | \mathrm{Tr} ( X \pi_\psi) \leq
  \mathcal{E}(|\psi\rangle) \text{ for } ^\forall|\psi \rangle \in \mathcal{H} \}.
  \nonumber
\end{align}
$\pi_\psi \equiv  | \psi \rangle \langle \psi |$ is the projector on a pure state $\psi$, $\mathrm{Tr}(\cdot)$ is the trace, and $\mathbb{M}_{\mathcal{H}}$ includes the null operator to have $\mathcal{E} \ge 0$.
A physical operator $X$ $\in \mathbb{M}_{\mathcal{H}}$ does not overestimate $\mathcal{E}(|\psi \rangle)$ for all pure states $\psi$ in a Hilbert space $\mathcal{H}$, hence $- X$ is a witness~\cite{Horodecki,Lewenstein,Terhal} detecting entanglement; see Fig.~\ref{fig1}(a).
Among $X$'s, the optimal one with the largest expectation value of $\mathrm{Tr}(X \rho)$ detects $\mathcal{E}(\rho)$, and a less optimal one provides a lower bound of $\mathcal{E}(\rho)$~\cite{Terhal,Brandao,Audenaert,Guhne_PRL,Eisert,Plenio}.
The optimal witness exists as a support hyperplane of a convex set of states [see Fig.~\ref{fig1}(b)], and is an observable in principle experimentally accessible~\cite{Park}.
Hereafter, we would call $X$ a witness, although $- X$ is a conventional one, and accordingly redefine the optimal one:

{\em Definition.}
The optimal witness $X_\rho \in \mathbb{M}_{\mathcal{H}}$
is defined relative to $\rho$ and $\mathcal{E}$.
It satisfies $\mathrm{Tr} (X_\rho \rho) = \mathcal{E}(\rho)$.

To compute $\mathcal{E}(\rho)$, one needs to find $X_\rho$. We below show the basic properties of $X_\rho$ 
useful for finding $X_\rho$ efficiently.

{\em Theorem 1.}
The optimal witness $X_\rho$ for $\rho$ is also optimal for all the pure states
$\psi^\mathrm{op}_i$ of the optimal decomposition of $\rho$ giving
$\mathcal{E}(\rho) = \sum_i p^\mathrm{op}_i \mathcal{E}(|\psi^\mathrm{op}_i \rangle)$.

\noindent{\it Proof by contradiction.}
Suppose that $X_\rho$ is not the optimal witness for $^\exists\psi^\mathrm{op}_j$,
$\mathrm{Tr}(X_\rho \pi_{\psi^\mathrm{op}_j}) < \mathcal{E}(\psi^\mathrm{op}_j)$.
Then, $\mathrm{Tr}(X_\rho \rho) = \sum_i p^\mathrm{op}_i \mathrm{Tr}(X_\rho \pi_{\psi^\mathrm{op}_i})
< \sum_i p^\mathrm{op}_i \mathcal{E}(|\psi^\mathrm{op}_i \rangle) = \mathcal{E} (\rho)$,
contradicting the fact that $X_\rho$ is optimal for $\rho$. $\ensuremath{\hfill \Box}$

{\em Theorem 2.}
Consider the set $\mathbb{P}_{X} \subset \mathcal{H}$ of the pure states $\{ | \psi_i \rangle \}$ for which a witness $X \in \mathbb{M}_{\mathcal{H}}$ is optimal, $\textrm{Tr} (X \pi_{\psi_i}) = \mathcal{E} (|\psi_i \rangle)$. Then, for ${\it any}$ convex mixture $\rho = \sum_i p_i \pi_{\psi_i}$ $\in \mathrm{conv}(\mathbb{P}_{X})$, $X$ is optimal, $X_\rho = X$, and $\sum_i p_i \pi_{\psi_i}$ is the optimal decomposition of $\rho$.

\noindent{\it Proof.}
$\mathcal{E}(\rho) \ge \mathrm{Tr}(X \rho) = \sum_i p_i \mathrm{Tr}(X \pi_{\psi_i})$ from Eq.~(2), 
$\mathcal{E} (\rho) \le \sum_i p_i \mathcal{E} (|\psi_i \rangle)$ from Eq.~(1),
and
$\mathrm{Tr} (X \pi_{\psi_i}) = \mathcal{E} (|\psi_i \rangle)$.
$\therefore \mathcal{E}(\rho) = \sum_i p_i \mathcal{E} (|\psi_i \rangle) = \textrm{Tr} (X \rho)$. $\ensuremath{\hfill \Box}$

The theorems connect $X_\rho$ and the optimal pure-state decomposition of $\rho$ in Eq.~\eqref{convroof}. They are valid for all convex measures.
Crucial are their consequences [Fig.~\ref{fig1}(c)]:
One can check the optimality of a witness $X$ for $\rho$, by seeing whether $\rho \in \mathrm{conv}(\mathbb{P}_{X})$.
Namely, one computes $\mathcal{E}(\rho)$ or its lower bound, by optimizing the form of $X_\rho$, or by guessing the form and checking its optimality for $\rho$ (with possibly avoiding heavy computation).
Moreover, with a single $X_\rho$, one can obtain the analytic expression of $\mathcal{E}$ or experimentally determine $\mathcal{E}$ for {\it all} $\rho \in \mathrm{conv}(\mathbb{P}_{X_\rho})$.
This considerably reduces not only computational but also experimental efforts.
For example, when a target state $\rho$ is affected by noise in experiments, the optimal witness $X_\rho$ of $\rho$ can still give the exact value of an entanglement measure of the affected state or a faithful lower bound, if the affected state is belonging to or close to $\mathrm{conv}(\mathbb{P}_{X_\rho})$.

We further provide the restrictions on the form of $X_\rho$ by the range $\mathcal{R}(\rho)$ and rank $r_\rho$ of $\rho$.
The Hilbert space $\mathcal{H}$ for $\mathbb{M}_\mathcal{H}$ is $\mathcal{R}(\rho)$ or the full Hilbert space $\mathcal{H}_\mathrm{f}$ of the system.
Theorem 1 ensures another restriction useful for large $r_\rho$:

{\em Corollary.} The number of linearly independent pure states in $\mathbb{P}_{X_\rho}$ should be larger than or equals to $r_\rho$.

Symmetries of $\mathcal{E}$ or of $\rho$ also restrict $X_\rho$.
Many useful measures
characterize SLOCC invariant entanglement. In this case, all pure states with finite $\mathcal{E}$ are connected, by SLOCC operations (tensor products of local operators with determinant 1), to a maximally entangled state.
This connection simplifies $X_\rho$; see, e.g., Eq.~\eqref{fullrankwit}.


Moreover, when $\rho$ has (higher) symmetries, it is enough to consider the (simpler) form of $X_\rho$ symmetrized by the same symmetries, as $\mathrm{Tr} (X_\rho \rho)$ indicates.
For highly symmetric states, our approach reproduces (hence covers) previous theoretical results~\cite{Terhal_PRL,Verstraete_4qubit,Wei04,Lohmayer}, and also enables experimental quantification, contrary to the previous works; for example, for an isotropic state and a bipartite Werner state, $X_\rho$ has only one parameter, hence one analytically obtains $X_\rho$ and $\mathcal{E}$ easily.
For states having low symmetry, multipartite, high rank, and/or nontrivial decomposition of Eq.~\eqref{convroof}, our approach is more useful for obtaining $X_\rho$ and $\mathcal{E}$ than the previous works; see below.



{\it Noisy Smolin states.--} For illustration, we first quantify entanglement in four-qubit noisy Smolin states
\begin{equation}
\rho_\textrm{NS} (p) = (1-p) \rho_\textrm{S} + \frac{pI}{16}, \,\,\,
\rho_\textrm{S} = \frac{1}{16}(I + \sum_{j=1}^3 \sigma_j^{\otimes 4}), \,\,\, p \in [0,1]
\nonumber
\end{equation}
by geometric measure of entanglement~\cite{Wei04} $\mathcal{E}_\mathrm{G}$;
$\sigma_j$'s are Pauli matrices and $I$ is the identity.
For $p \in [0, 2/3)$, $\rho_\textrm{NS} (p)$ has bound entanglement experimentally more reliable than Smolin state $\rho_\mathrm{S}$~\cite{Smolin,Amselem}. It was realized experimentally~\cite{Lavoie}.
Its entanglement has never been quantified, 
while $\mathcal{E}_\mathrm{G}(\rho_\textrm{S})$ was computed in Ref~\cite{Wei04}.

We derive a general condition of $X$ for $\mathcal{E}_\mathrm{G}$, which greatly reduces computation costs. With the set $\mathbb{U}$ of separable pure states, $\mathcal{E}_\mathrm{G}$ is defined~\cite{Wei04} as $\mathcal{E}_\mathrm{G} ( | \psi \rangle ) = 1 - \max_{| s \rangle \in \mathbb{U}} | \langle \psi | s \rangle |^2$ for pure states $\psi$, and extended to mixed cases via Eq.~\eqref{convroof}. Since $\langle \psi |X| \psi \rangle \le \mathcal{E}_\mathrm{G} (|\psi\rangle) \le 1 - |\langle \psi | s \rangle|^2$, $\langle \psi | X + \pi_s | \psi \rangle \le 1$ for $^\forall |s \rangle \in \mathbb{U}$ and $^\forall X$. The equality holds, when $X$ is optimal for $\psi$ and $|s\rangle$ is the state in $\mathbb{U}$ with maximal $|\langle \psi | s \rangle|^2$. Since any $X$ is optimal for at least one pure state (see Corollary), it satisfies
\begin{equation}
\max_{| s \rangle \in \mathbb{U}} \lambda_{\max} [ X + \pi_s ] = 1,
\label{wit_geo}
\end{equation}
where $\lambda_{\max} [A]$ denotes the maximum eigenvalue of $A$.


We now turn to $\rho_\textrm{NS} (p)$. Its symmetries restrict $X_{\rho_\textrm{NS}}(p)$ to the form of $X(p) = 4 \alpha(p) \rho_\textrm{S} + \beta(p) (I - 4 \rho_\textrm{S})$ with only two real parameters $\alpha \geq 0$ and $\beta \leq 0$. From the witness condition of Eq.~\eqref{wit_geo}, one has $\alpha^{-1} + \beta^{-1} = 2$ or $\alpha = \beta = 0$. By maximizing $\mathrm{Tr} ( X \rho_\mathrm{NS} (p) ) = [(4-3p)\alpha + 3p\beta]/4$ over $\alpha$ and $\beta$, we obtain for $p \in [0, 2/3)$
\begin{equation}
\mathcal{E}_\mathrm{G} (\rho_\mathrm{NS}(p)) = [ 2 - \sqrt{3 p(4-3p)} ] / 4, 
\end{equation}
$\alpha (p) = [ 1 - \sqrt{\frac{3p}{4-3p}} ] /2$, and $\beta (p) = [ 1 - \sqrt{\frac{4-3p}{3p}} ] /2$, while $\mathcal{E}_\mathrm{G} (\rho_\mathrm{NS}(p)) = \alpha (p) = \beta (p) = 0$ for  $p \in [2/3, 1]$.
The analytic result of $\mathcal{E}_\mathrm{G} (\rho_\mathrm{NS})$ agrees with previous findings of separability for $p \in [2/3, 1]$~\cite{Augusiak} and $\mathcal{E}_\mathrm{G} (\rho_\mathrm{S}) = 1/2$~\cite{Wei04}, and does not require heavy optimization, which might be necessary in
previous approaches~\cite{Terhal_PRL,Wei04}.


$X_{\rho_\textrm{NS}}$ is accessible in experiments.
The construction of the common component of $X_{\rho_\textrm{NS}}$, $\sum_j \sigma_j^{\otimes 4}$, which was done in Ref.~\cite{Lavoie}, provides all $X_{\rho_\textrm{NS}}(p)$'s.

{\it 3-qubit GHZ entanglement.--} We next consider three-qubit GHZ entanglement. Using its SLOCC invariance, we derive the general form of the optimal witness.


In three qubits, there are two types of genuine tripartite entanglement, GHZ and W~\cite{Dur,Bennett,Acin}.
Their representative states are
$|\mathrm{GHZ} \rangle = \frac{1}{\sqrt{2}} ( |000\rangle + |111\rangle )$
and $|\mathrm{W} \rangle = \frac{1}{\sqrt{3}} ( |100\rangle + |010\rangle + |001\rangle )$.
To quantify GHZ entanglement, we define a new entanglement measure $\mathcal{T}_3$~\cite{Park}
by $\mathcal{T}_3 (| \psi \rangle) = \sqrt{ \tau_3 (| \psi \rangle )}$ for pure states, and extend it to mixed states via Eq.~\eqref{convroof}, where $\tau_3$ is three-tangle~\cite{Coffman}.
It is ``extensive'',
$\mathcal{T}_3 (|\psi \rangle) = \langle \psi | \psi \rangle \;
\mathcal{T}_3 ( |\psi \rangle/ \sqrt{\langle \psi | \psi \rangle} )$, resulting in
the property useful for $X_\rho$ that $\mathcal{T}_3$ is SLOCC invariant for pure and mixed states, $\mathcal{T}_3(\rho) = \mathcal{T}_3(O \rho O^\dagger)$.
Any pure state in GHZ$\setminus$W class is transformed into $|\mathrm{GHZ} \rangle$,
\begin{equation}
  \label{normform}
  |\psi \rangle = \sqrt[4]{\tau_3 (|\psi \rangle)} \; O_\psi |\mathrm{GHZ} \rangle = \sqrt{\mathcal{T}_3 (|\psi \rangle)} \; O_\psi |\mathrm{GHZ} \rangle,
\end{equation}
by an SLOCC operator $O_\psi$~\cite{Verstraete}.
We call $\mathcal{T}_3$ {\it extensive} three-tangle, and use it instead of three-tangle $\tau_3$, since $\tau_3$ is not SLOCC invariant for mixed states~\cite{Supp}.

We demonstrate how to construct $X_\rho$ for $\mathcal{T}_3$.
Theorem 1 ensures that $X_\rho$
should be optimal also for at least one pure state $\psi$ in GHZ$\setminus$W class,
$\mathrm{Tr}(X_\rho |\psi \rangle \langle \psi |) = \mathcal{T}_3 (| \psi \rangle)$.
This property and Eq.~\eqref{normform} result in
$O^\dagger_\psi X_\rho O_\psi |\mathrm{GHZ} \rangle = | \mathrm{GHZ} \rangle$,
which gives the general form of $X_\rho$~\cite{Supp},
\begin{equation}
  \label{fullrankwit}
    X_\rho = O \;( \pi_{\mathrm{GHZ}} + \Pi - \mu I ) \; O^\dagger / (1 - \mu),
    \,\,\,\,\, O \in \mathrm{SLOCC},
\end{equation}
where $\pi_\mathrm{GHZ} \equiv | \mathrm{GHZ} \rangle \langle \mathrm{GHZ} |$. 
It generalizes the widely-used witness of $\pi_\mathrm{GHZ} - 3I/4$~\cite{Acin}, and indicates that for $\rho$ with finite $\mathcal{T}_3$, $\rho$ is optimally decomposed into one GHZ$\setminus$W-class state and other W-class states. $\mu = \max_{\varphi \in \mathrm{W}} \langle \varphi | \pi_{\mathrm{GHZ}} + \Pi | \varphi \rangle$
prevents $X_\rho$ from overestimating $\mathcal{T}_3$ for all pure states in $\mathcal{H}$. To obtain $\mu$, it is enough to consider pure states $\varphi$ (with $\langle \varphi | \varphi \rangle = 1$) in W class~\cite{Supp}.

$\Pi$ is a sum of pure-state projectors
satisfying $0 \le \Pi < \mu I$ and $\mathrm{Tr} (\Pi \pi_\mathrm{GHZ}) = 0$.
For states with $r_\rho > 2$, only specific forms of $\Pi$ satisfy the Corollary.
For example, $\Pi = \lambda \pi_\mathrm{W} = \lambda | \mathrm{W} \rangle \langle \mathrm{W} |$
cannot be used for $r_\rho > 4$, since $\mathbb{P}_{X_\rho}$ contains only four states.
Our numerical search~\cite{Supp} might imply that the Corollary is satisfied only if
$\Pi$ is invariant under some of the symmetries of $\pi_\textrm{GHZ}$ and $I$ such as permutation, exchange of qubit index, 0-1 flip, local phase rotation of
$U (\alpha, \beta) =
      \left(
      \begin{smallmatrix}
        1 & 0 \\
        0 & e^{i\alpha}
      \end{smallmatrix}
      \right)
      \otimes
      \left(
      \begin{smallmatrix}
        1 & 0 \\
        0 & e^{i\beta}
      \end{smallmatrix}
      \right)
      \otimes
      \left(
      \begin{smallmatrix}
        1 & 0 \\
        0 & e^{-i(\alpha+\beta)}
      \end{smallmatrix}
      \right)$ with real $\alpha$ and $\beta$, etc.
It will be valuable to prove this conjecture.
A more symmetric form of $\Pi$ has bigger $\mathbb{P}_{X_\rho}$ and smaller number of parameters.
For instance, $\Pi_\mathrm{S} \equiv \lambda_{\overline{\mathrm{GHZ}}} \pi_{\overline{\mathrm{GHZ}}} + \sum_{ijk \neq 000, 111} \lambda_{ijk} \pi_{ijk}$ has the biggest set of $\mathbb{P}_{X_\rho}$, as it is invariant under all available local phase rotations, where
$|\overline{\mathrm{GHZ}} \rangle = (|000 \rangle - |111 \rangle)/\sqrt{2}$ and
$\pi_{ijk} = |ijk \rangle \langle ijk|$. Symmetric forms such as $\Pi_\textrm{S}$ are useful for the full-rank cases of $r_\rho =8$.

We discuss the efficiency of our strategy for $\mathcal{T}_3$.
For a simple form of $\rho$, one guesses a trial witness form $X$, optimizes it, checks the optimality whether $\rho \in \mathrm{conv}(\mathbb{P}_{X})$, then computes $\mathcal{E}(\rho) = \textrm{Tr} (X \rho)$.
This procedure is useful for $\rho$ with symmetries, for which one chooses $X$ with the same symmetries.
On the other hand, for a complex form of $\rho$, one fully optimizes the form of $X_\rho$ in Eq.~\eqref{fullrankwit}.
This optimization still has a cost much cheaper than the direct pure-state decomposition of Eq.~\eqref{convroof}.
For $r_\rho = 8$, $X_\rho$ has 72 optimization parameters (48 for deciding the eigenstates of $\Pi$, 6 for the eigenvalues of $\Pi$, and $18$ for $O$), while the direct decomposition has hundreds to a thousand (roughly $2 r_\rho^3$) of parameters~\cite{Roth_PRL}.
If the conjecture about the symmetric forms of $\Pi$ is true, the parameter number of $X_\rho$ is further reduced to at most 40; $X_\rho$ with $\Pi_\textrm{S}$ has 24 parameters. Even in the case that the conjecture is false, a symmetric $\Pi$ is useful, as it gives large $\mathbb{P}_{X_\rho}$ and at least a good lower bound of $\mathcal{T}_3$.

\begin{figure}[t]
  \includegraphics[width=.42\textwidth]{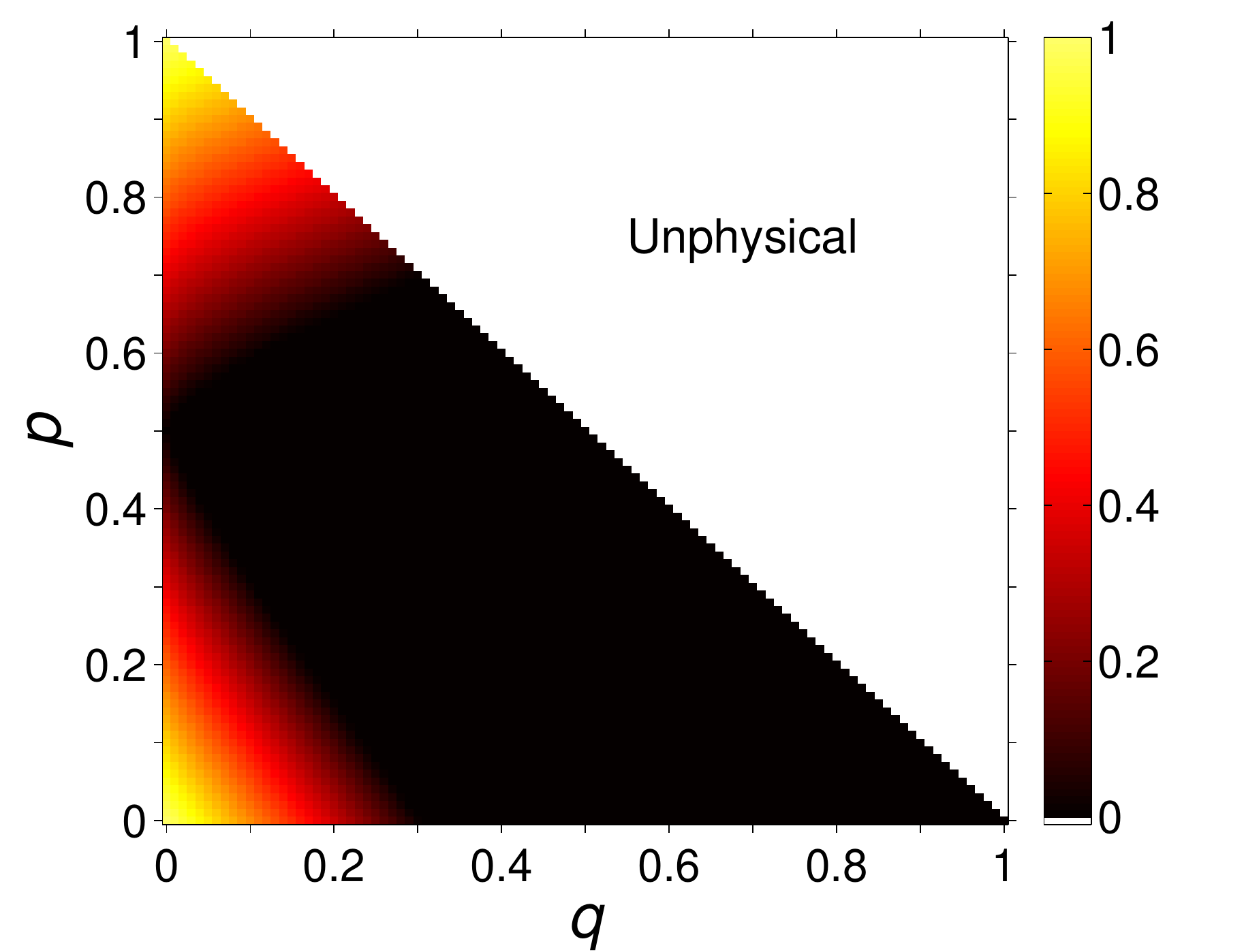}
  \caption{(Color Online) {Three-qubit mixed-state GHZ entanglement.} The exact value of GHZ entanglement measure $\mathcal{T}_3$ is computed for the 3-qubit full-rank state $\rho_\mathrm{G\overline{G}I}(p,q)$, by using the optimal witness operator $X_{\rho_\mathrm{G\overline{G}I}}$ for $\rho_\mathrm{G\overline{G}I}(p,q)$.
  The physical region of $(p,q)$ is defined by $0 \le p, q, (1-p-q) \le 1$.}
  \label{fig2}
\end{figure}

We provide examples.
We construct $X_\rho$ for arbitrary mixtures of $|\mathrm{GHZ} \rangle$,
$|\overline{\mathrm{GHZ}} \rangle$ and the white noise $I/8$, $\rho_\mathrm{G\overline{G}I} (p,q)
= (1-p-q) \pi_\mathrm{GHZ} + p \pi_{\overline{\mathrm{GHZ}}} + q I/8$,
\begin{eqnarray}
  X_{\rho_\mathrm{G\overline{G}I}} & = & ( \pi_\mathrm{GHZ} + \lambda \pi_{\overline{\mathrm{GHZ}}} - \mu I)/ (1 - \mu).
    \label{ghzimixwit}
\end{eqnarray}
Here, the symmetries of $\rho_\mathrm{G\overline{G}I}$ fix
$\Pi = \lambda \pi_{\overline{\mathrm{GHZ}}}$.
Checking the optimality, we fix $\lambda$, and compute $\mathcal{T}_3$ in Fig.~\ref{fig2}.

For any mixture $\rho_\mathrm{GI}(q) [\equiv \rho_\mathrm{G\overline{G}I}(p=0,q)]$
of $|\mathrm{GHZ} \rangle$ and $I/8$ with $q \le q_0 \simeq 0.304$, $X_{\rho_\mathrm{GI}} \equiv
X_{\rho_\mathrm{G\overline{G}I}}(\mu \simeq 0.750, \lambda \simeq 0.433)$ is optimal,
independent of $q$. The result of $\mathcal{T}_3$ is
\begin{equation}
  \label{ghzimix}
  \mathcal{T}_3 = \max [0,\mathrm{Tr} (X_{\rho_\mathrm{GI}} \rho_\mathrm{GI})] =
    1- q / q_0 \,\,\, \textrm{for } q \leq q_0,
\end{equation}
$\mathcal{T}_3 = 0$ for $q > q_0 \simeq 0.304$.
The optimal decomposition is
$\rho_\mathrm{GI}(q) = \mathcal{T}_3 (q) \pi_\mathrm{GHZ} + (1-\mathcal{T}_3(q)) \rho_\mathrm{Z}$ for  $q \leq q_0$, and
$ \rho_\mathrm{GI}(q) = \frac{1-q}{1-q_0} \rho_\mathrm{Z} + \frac{q-q_0}{1-q_0} \frac{I}{8}$ for $q > q_0$, where
$\rho_\mathrm{Z} = \; (1-q_0) \pi_\mathrm{GHZ} + q_0 I/8 \propto \sum_i \pi_{\mathrm{Z}_i}$
is decomposed by W-class states $|\mathrm{Z}_i \rangle$ with $\mathrm{Tr}(X_{\rho_\mathrm{GI}} \pi_{\mathrm{Z}_i})=0$; $|Z_i \rangle$'s are given in Ref.~\cite{Supp}.
This computation has a cost significantly cheaper than the direct decomposition of Eq.~\eqref{convroof} which has 239 optimization parameters for $\rho_\textrm{GI}$.


We consider another state, the rank-2 mixture
$\rho_\mathrm{GW} (p) = (1-p) \pi_\mathrm{GHZ} + p \pi_\mathrm{W}$, $0 \le p \le 1$.
$X_\rho$ depends on the choice of $\mathcal{H}$ between $\mathcal{R}(\rho_\mathrm{GW})$ and $\mathcal{H}_\mathrm{f}$ (the full space).
For $\mathcal{H} = \mathcal{R}(\rho_\mathrm{GW})$,
a single witness quantifies $\mathcal{T}_3(\rho_\mathrm{GW})$,
\begin{eqnarray}
  X_{\rho_\mathrm{GW}}^{(1)} & = & \pi_\mathrm{GHZ} - (p_0^{-1} - 1) \pi_\mathrm{W},
  \label{Wwitness} \\
  \mathcal{T}_3 & = & \max [0, \mathrm{Tr}(X_{\rho_\mathrm{GW}}^{(1)}
  \rho_\mathrm{GW})] = 1 - p/p_0 \,\,\, \textrm{for } p \leq p_0,
  \nonumber
\end{eqnarray}
$ \mathcal{T}_3 = 0$ for $p > p_0 \simeq 0.373$; the optimal decomposition of $\rho_\mathrm{GW} (p)$ is given in Ref.~\cite{Supp}.
For $\mathcal{H} = \mathcal{H}_\mathrm{f}$,
$ X_{\rho_\mathrm{GW}}^{(2)} = \lim_{\lambda \rightarrow 1} \frac{1}{1- \mu}
(\pi_\mathrm{GHZ} + \lambda \pi_\mathrm{W} - \mu I)$
gives $\mathcal{T}_3(\rho_\mathrm{GW})$
asymptotically with $\lim_{\lambda \to 1}$.

The examples are useful for understanding the fragility of GHZ entanglement under noise.
$\rho_\mathrm{G\overline{G}I}$
includes the white noise and the dephasing of the relative phase between $|000 \rangle$ and $|111 \rangle$.
GHZ entanglement in $\rho = (1-\alpha) \pi_\textrm{GHZ} + \alpha \Pi'$
decreases as $\alpha$ increases, and vanishes
at $\alpha = 0.5$ for $\Pi' = \pi_{\overline{\mathrm{GHZ}}}$, 0.373
for $\pi_\mathrm{W}$, and 0.304 for $I/8$;
for $\pi_{\overline{\mathrm{GHZ}}}$, GHZ entanglement revives and increases
with $\alpha > 0.5$.
The white noise destroys GHZ entanglement more than $\pi_{\overline{\mathrm{GHZ}}}$ and
$\pi_\mathrm{W}$, and more severely for states with more qubits;
$\alpha = 0.304$ is smaller than $\beta = 2/3$ at which Bell-state ($|\textrm{Bell}\rangle$) entanglement in $\rho_2 = (1-\beta) |\textrm{Bell} \rangle \langle \textrm{Bell}| + \beta I/ 4$ vanishes.

From our finding, one can quantify GHZ entanglement in experiments. When $|\mathrm{GHZ} \rangle$ is prepared, it normally becomes $\rho_\mathrm{G\overline{G}I}$ due to noise.
Assuming that the prepared state $\rho_\textrm{exp}$ has the form of $\rho_\mathrm{G\overline{G}I}(p,q)$,
one estimates $\mathcal{T}_3$ by $X_{\rho_\mathrm{G\overline{G}I}}$:
One measures $\pi_\mathrm{GHZ}$ and $\pi_{\overline{\mathrm{GHZ}}}$~\cite{Guhne}, and
computes the largest value of $\textrm{Tr} (X_{\rho_\mathrm{G\overline{G}I}} \rho_\textrm{exp})$
with varying $(\mu,\lambda)$
over the space where $X_{\rho_\mathrm{G\overline{G}I}}$ is a witness.
Then, it is the exact value (a faithful lower bound) of $\mathcal{T}_3$
if the assumption is correct (incorrect).
This procedure is powerful, as $\mathcal{T}_3$ is obtained from minimal information about $\rho_\textrm{exp}$.

{\it Conclusion.--} Our approach of optimal witness has great advantage over previous methods of optimizing state decomposition. It offers a simple way of theoretical and experimental quantification of entanglement prepared in laboratories (which is usually in a simple state such as $\rho_\mathrm{NS}$ and $\rho_\mathrm{G\overline{G}I}$), and stimulates researches on multipartite or high-dimensional mixed-state entanglement.




We thank O. G\"{u}hne, M. B. Plenio, and T.-C. Wei for useful discussion, and
NRF for support (2009-0084606).


\begin{thebibliography}{99}



\bibitem{Plenio_review} M. B. Plenio and S. Virmani,
Quant. Inf. Comp. {\bf 7}, 1 (2007); R. Horodecki, P. Horodecki, M. Horodecki, and K. Horodecki,
Rev. Mod. Phys. {\bf 81}, 865 (2009).

\bibitem{Guhne} O. G\"{u}hne and G. T\'{o}th,
Phys. Rep. {\bf 474}, 1 (2009).

\bibitem{Amico}
L. Amico, R. Fazio, A. Osterloh, and V. Vedral,
Rev. Mod. Phys. {\bf 80}, 517 (2008).

\bibitem{Wootters} W. K. Wootters,
Phys. Rev. Lett. {\bf 80}, 2245 (1998).

\bibitem{Schmid} C. Schmid {\it et al.},
Phys. Rev. Lett. {\bf 101}, 260505 (2008); F. Mintert and A. Buchleitner, {\it ibid.} {\bf 98}, 140505 (2007).


\bibitem{Park} H. S. Park, S.-S. B. Lee, H. Kim, S.-K. Choi, and H.-S. Sim, Phys. Rev. Lett {\bf 105}, 230404 (2010); S.-S. B. Lee and H.-S. Sim, Phys. Rev. A {\bf 79}, 052336 (2009).

\bibitem{Terhal_PRL} B. M. Terhal and K. G. H. Vollbrecht,
Phys. Rev. Lett. {\bf 85}, 2625 (2000); K. G. H. Vollbrecht and R. F. Werner,
Phys. Rev. A {\bf 64}, 062307 (2001).

\bibitem{Verstraete_4qubit} F. Verstraete, J. Dehaene, B. De Moor, and H. Verschelde,
Phys. Rev. A {\bf 65}, 052112 (2002).

\bibitem{Wei04} T.-C. Wei, J. B. Altepeter, P. M. Goldbart, and W. J. Munro,
Phys. Rev. A {\bf 70}, 022322 (2004).

\bibitem{Lohmayer} R. Lohmayer, A. Osterloh, J. Siewert, and A. Uhlmann,
Phys. Rev. Lett. {\bf 97}, 260502 (2006).


\bibitem{Brandao} F. G. S. L. Brand\~{a}o,
Phys. Rev. A {\bf 72}, 022310 (2005).

\bibitem{Terhal} B. M. Terhal,
{\it Theor. Comput. Sci.} {\bf 287}, 313-335 (2002).

\bibitem{Lewenstein} M. Lewenstein, B. Kraus, J.I. Cirac, and P. Horodecki,
Phys. Rev. A {\bf 62}, 052310 (2000).

\bibitem{Augusiak} R. Augusiak and P. Horodecki,
Phys. Rev. A {\bf 74}, 010305(R) (2006).

\bibitem{Lavoie} J. Lavoie, R. Kaltenbaek, M. Piani, and K. J. Resch,
Phys. Rev. Lett. {\bf 105}, 130501 (2010).


\bibitem{GHZ}
D. M. Greenberger, M. A. Horne, A. Schimony, and A. Zeilinger,
Am. J. Phys. {\bf 58}, 1131 (1990).

\bibitem{Eisert} J. Eisert, F. G. S. L. Brand\~{a}o, and K. M. R. Audenaert,
New J. Phys. {\bf 9}, 46 (2007).

\bibitem{Horodecki} M. Horodecki, P. Horodecki, and R. Horodecki,
Phys. Lett. A {\bf 223}, 1 (1996).

\bibitem{Audenaert} K. M. R. Audenaert and M. B. Plenio, New J. Phys. {\bf 8}, 266 (2006).

\bibitem{Guhne_PRL} O. G\"{u}hne, M. Reimpell, and R. F. Werner,
Phys. Rev. Lett. {\bf 98}, 110502 (2007).

\bibitem{Plenio} M. B. Plenio, Science {\bf 324}, 342 (2009).

\bibitem{Smolin} J. A. Smolin,
Phys. Rev. A {\bf 63}, 032306 (2001).

\bibitem{Amselem} E. Amselem and M. Bourennane,
Nat. Phys. {\bf 5}, 748 (2009).

\bibitem{Dur} W. D\"{u}r, G. Vidal, and J. I. Cirac,
Phys. Rev. A {\bf 62}, 062314 (2000).

\bibitem{Bennett} C. H. Bennett {\it et al.},
Phys. Rev. A {\bf 63}, 012307 (2000).

\bibitem{Acin} A. Ac\'{i}n, D. Bru{\ss}, M. Lewenstein, and A. Sanpera,
Phys. Rev. Lett. {\bf 87}, 040401 (2001).

\bibitem{Coffman} V. Coffman, J. Kundu, and W. K. Wootters,
Phys. Rev. A {\bf 61}, 052306 (2000).

\bibitem{Verstraete} F. Verstraete, J. Dehaene, and B. De Moor,
Phys. Rev. A {\bf 68}, 012103 (2003).

\bibitem{Roth_PRL} B. R\"{o}thlisberger {\it et al.},
Phys. Rev. Lett. {\bf 100}, 100502 (2008).

\bibitem{Supp} See the Supplementary Information.

\end{thebibliography}
\end{document}